\documentclass[12pt,preprint]{aastex}

\usepackage{aas_macros,graphicx,amsmath}
\defcitealias{2009MNRAS.400.1121S}{S09}

\shorttitle{Observational limits on the gas mass of a $\lowercase{z} = 4.9$ galaxy}
\shortauthors{R.~C.~Livermore et~al.}

\begin{document}


\title{Observational limits on the gas mass of a $\lowercase{z}=4.9$ galaxy}


\author{R.~C.~Livermore\altaffilmark{1}, A.~M.~Swinbank\altaffilmark{1}, Ian Smail\altaffilmark{1}, R.~G.~Bower\altaffilmark{1}, K.~E.~K. Coppin\altaffilmark{1,2}, R.~A. ~Crain\altaffilmark{3}, A.~C. Edge\altaffilmark{1}, J.~E.~Geach\altaffilmark{1,2}, and J.~Richard\altaffilmark{1,4}}


\altaffiltext{1}{Institute for Computational Cosmology, Durham University, South Road, Durham DH1 3LE, UK}
\altaffiltext{2}{Department of Physics, McGill University, 3600 rue University, Montr\'eal, QC H3A 2T8, Canada}
\altaffiltext{3}{Leiden Observatory, Leiden University, PO Box 9513, 2300 RA Leiden, The Netherlands}
\altaffiltext{4}{CRAL, Observatoire de Lyon, 9 Avenue Charles Andr\'e, 69561 Saint-Genis-Laval, France}


\begin{abstract}
We present the results of a search for molecular gas emission from a star-forming galaxy at $z = 4.9$. The galaxy benefits from magnification of $22 \pm 5\times$ due to strong gravitational lensing by the foreground cluster MS1358+62. We target the CO(5--4) emission at a known position and redshift from existing \emph{Hubble Space Telescope}/\emph{ACS} imaging and Gemini/NIFS [O{\sc ii}]3727 imaging spectroscopy, and obtain a tentative detection at the $4.3\,\sigma$ level with a flux of $0.104 \pm 0.024$\,Jy\,km\,s$^{-1}$. From the CO line luminosity and assuming a CO-to-H$_2$ conversion factor $\alpha=2$, we derive a gas mass $M_{\rm{gas}} \sim 1^{+1}_{-0.6} \times 10^9$\,M$_{\sun}$. Combined with the existing data, we derive a gas fraction $M_{\rm{gas}}/\left(M_{\rm{gas}} + M_{\ast}\right) = 0.59^{+0.11}_{-0.06}$. The faint line flux of this galaxy highlights the difficulty of observing molecular gas in representative galaxies at this epoch, and suggests that routine detections of similar galaxies in the absence of gravitational lensing will remain challenging even with ALMA in full science operations.
\end{abstract}





\section{Introduction}
\label{sec:intro}

Extensive studies of optically-selected star-forming galaxies seen at the
epoch of peak cosmic star-formation density ($z \sim 2$) have revealed
star-formation rates (SFRs) of 10-100\,M$_{\odot}$yr$^{-1}$ and stellar
masses of $\sim$10$^{10-11}$M$_{\odot}$
\citep[e.g.][]{2009ApJ...706.1364F,2009ApJ...697.2057L,2010Natur.463..781T,2010ApJ...713..686D}. If
these SFRs have been continously maintained, then these galaxies must
have undergone their first major epoch of stellar mass assembly
1--2\,Gyr earlier, at $z\sim 5$, when the bulk of the
star-forming population was $\sim 5\times$ less massive
\citep{2009MNRAS.395.2196M,2010MNRAS.408.1628S}.

The star formation within these galaxies is fuelled by reservoirs of
predominantly cold molecular hydrogen, H$_{2}$. Since the H$_{2}$ is
not directly detectable, CO emission at millimetre wavelengths has been
employed to trace the cold molecular gas. However, exploring gas
properties of galaxies beyond $z\sim 3$, is challenging; not only does
the apparent surface brightness reduce as $(1 + z)^{-4}$, but the
galaxies themselves also appear systematically smaller and
intrinsically fainter making detections of their molecular gas emission
difficult.

To date, studies of molecular line emission at $z > 3$ have been limited to extreme populations such
as gas-rich quasars \citep[e.g.][]{2009ApJ...691L...1W} and
submillimetre galaxies
\citep[SMGs;][]{2010MNRAS.407L.103C,2010ApJ...714.1407C,2011ApJ...739L..31R}. Detecting typical star-forming galaxies at $z>3$ has
proven difficult because the CO line luminosities are usually below
the sensitivity limits of current facilities
\citep{2008ApJ...687L...1S,2010MNRAS.408L..31D}.

However, it is still possible to study
less active high-$z$ galaxies which are strongly
gravitationally lensed by massive galaxy clusters. The magnification
provided by gravitational lensing has enabled detections of molecular
gas in a number of star-forming galaxies up to $z \sim 3$
\citep[e.g.][]{2004ApJ...604..125B,2007ApJ...665..936C,2011ApJ...733L..12R,2011MNRAS.410.1687D},
and more recently $z \sim 4-6$
\citep[e.g.][]{2011ApJ...740...63C,2012A&A...538L...4C}. The physical
properties of the interstellar medium appear similar to those in local
ULIRGs, with high gas fractions, high densities and intense UV
radiation fields.

\citet{1997ApJ...486L..75F} reported the detection of a multiply-imaged
$z=4.9$ galaxy which is gravitationally lensed by the massive galaxy cluster MS\,1358+62. Correcting
for lensing, this galaxy (hereafter MS1358-arc) appears to be
representative of the star-forming population at this epoch (its
lensing-corrected apparent magnitude is I$_{AB}$=24.9, marginally brighter than the characteristic luminosity of I$_{AB}=25.3$
at $z \sim 5$; \citealt{2004ApJ...611..660O}). \citet[hereafter
S09]{2009MNRAS.400.1121S} carried out a detailed study of one image of
the MS1358-arc system, using optical and infrared imaging combined
with integral field spectroscopy, revealing a rotating system across 2\,kpc in projection, with star formation occuring in five
bright clumps. 

In this Letter, we report observations with the Plateau de Bure
Interferometer to search for CO(5-4) emission in
MS1358-arc. Throughout, we adopt a $\Lambda$CDM cosmology with $H_0 =
70$\,km\,s$^{-1}$\,Mpc$^{-1}$, $\Omega_{\Lambda} = 0.7$ and $\Omega_m =
0.3$.

\section{Observations and Data Reduction}

\begin{figure*}
\includegraphics[width=174mm]{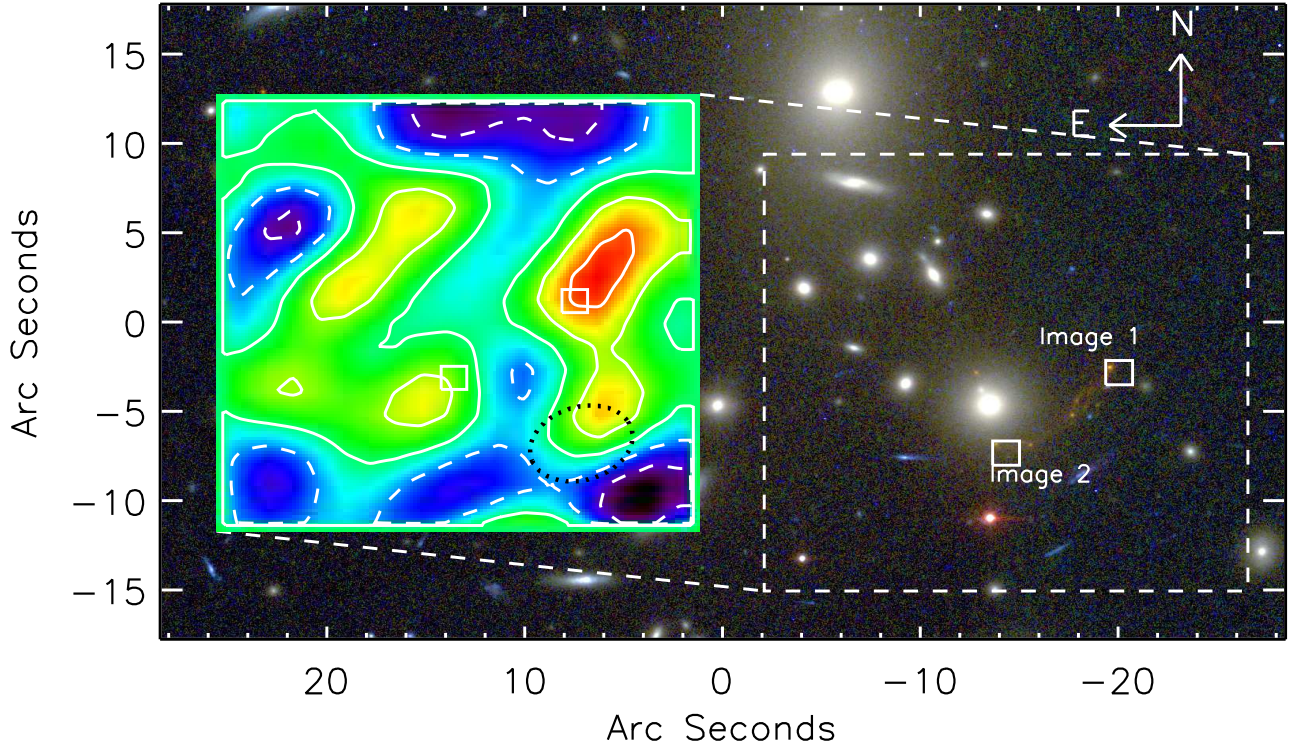}
\caption{\emph{Hubble Space Telescope (HST)/ACS} image of the core of MS1358+62.  The squares mark the extent of the pixels in the PdBI cube used to extract the spectra shown in Figure \ref{fig:specs}. A Plateau de Bure channel map constructed from a
  90km\,s$^{-1}$ window around the systemic velocity of the CO(5-4)
  line is shown as an inset. The PdBI synthesised beam
  size is indicated by the dotted ellipse in the inset. Contours are shown in $\pm 1\sigma$ intervals, with negative contours indicated by dashed lines. The origin of the {\it HST} image coordinate system is $(\alpha_{2000},\delta_{2000}) = (13^h 59^m 51.4^s, +62\degr
  30\arcmin 52.5\arcsec)$.}
\label{fig:hstimg}
\end{figure*}

The MS1358+62 system is illustrated in Figure \ref{fig:hstimg}, with the
positions of the two images of MS1358-arc marked (Table \ref{tab:ims}). We estimate the
amplification of each image using the lens model of
\citet{2008ApJ...685..705R} with \verb'lenstool'
\citep{1993PhDT.......189K,2007NJPh....9..447J}, with errors accounting
for the magnification gradient across the image. The combined
magnification factor for Image 1 and 2 is $\mu = 22 \pm 5$.

To search for the CO(5--4) emission, we observed MS1358-arc
with the IRAM Plateau de Bure Interferometer
\citep[PdBI;][]{1992A&A...262..624G}, using six antennae in D (compact)
configuration. The pointing centre was $\alpha_{2000}, \delta_{2000} =
13:59:48.7, +62:30:48.34$, and the frequency was tuned to the CO(5--4)
transition (rest frame 576.2679\,GHz) redshifted to 97.185\,GHz (based
on the [O{\sc ii}]-derived systemic redshift of $z=4.9296 \pm 0.0002$ from
\citetalias{2009MNRAS.400.1121S}). The observations were made between
2010 June 4 and 2010 June 6 with total on-source time of 10 hours,
using {\sc WIDEX} with a resolution of 2.5\,MHz. The star MWC349 was
observed as the primary flux calibrator, with the quasars 1749+096,
0923+392 and 2145+067 as secondary flux calibrators. Receiver bandpass
calibration was performed against 0923+392 and 3C454.3, and 1435+638
and 1418+546 were used for phase and amplitude calibration.

The data were reduced with the \verb'gildas' software package
\citep{2000ASPC..217..299G} and resampled to a velocity resolution of
38.6\,km\,s$^{-1}$. The synthesised beam is a Gaussian ellipse with a FWHM size of $5.6 \times 4.3
\arcsec$ at a position angle of $111\degr$.

The PdBI observations were centered on the two brightest images of the $z \sim 5$ galaxy. The images are both within 6'' of the pointing centre, whereas the primary beam of PdBI at 3mm is 50''; we have therefore not corrected for primary beam attenuation, which is negligible. First, we construct a channel map from a 90km\,s$^{-1}$ window around the CO(5--4) emission line centered at the systemic redshift, and show as an inset in Figure \ref{fig:hstimg}, with contours at $\pm 1\sigma$ intervals. The positions of the images within the PdBI cube are determined by aligning the cube with the \emph{HST} image and taking the centre of the corresponding pixels.

\begin{figure*}
\includegraphics[width=174mm]{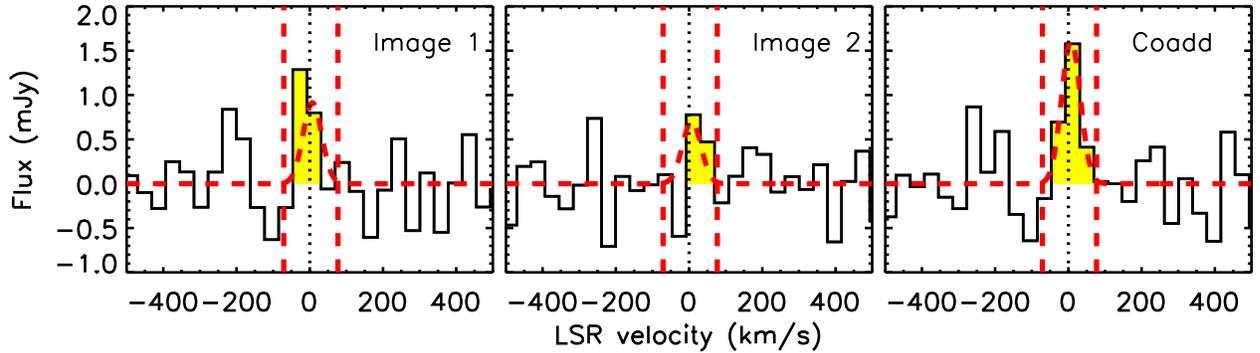}
\caption{The CO(5--4) spectra extracted from the two images of MS1358-arc and the coadd of these spectra. The velocity scale is centered on the systemic redshift $z=4.9296$, and the expected position of the CO(5--4) emission line is indicated by a black dotted line. The spectra are binned into
  38.6\,km\,s$^{-1}$ channels. The best-fit Gaussian profile to the coadded spectrum is overlaid as a
  guide, scaled to the relative amplifications of the two images, and the velocities of the two brightest star-forming clumps measured by \citetalias{2009MNRAS.400.1121S} are indicated by vertical lines. Each image shows positive flux at the systemic
  redshift of MS1358-arc, but individually they are only detected at $\sim
  3\sigma$ significance.}
\label{fig:specs}
\end{figure*}

We also extract spectra from these positions and show these in Figure \ref{fig:specs}. Individually, the two images are clearly extremely faint; by comparing the $\Delta \chi^2$ of a Gaussian profile to that of a continuum-only fit, measuring the noise over a 1200\,km\,s$^{-1}$ channel, the significances of the CO(5--4) emission lines in Images 1 and 2 are 3.3$\sigma$ and 2.7$\sigma$ respectively. However, since they are separated by more than one beam width, we can improve the signal to noise by coadding the spectra at the two positions using uniform weights (Figure \ref{fig:specs}); we then obtain a $1\sigma$ rms channel sensitivity of 0.4\,mJy per 38.6\,km\,s$^{-1}$ channel. Again comparing the $\chi^2$ of a Gaussian profile fit to that of a continuum-only fit, we use the $\Delta \chi^2$ to derive a significance level of 4.3\,$\sigma$ for the coadded CO(5--4) emission line.

\begin{table}
\caption{Positions of the three images of MS1358-arc within the Plateau
  de Bure field of view and the mean linear magnification factor $(\mu)$. The position of the third image is included for completeness, but is excluded from our analysis as it lies outside the primary beam.}
\label{tab:ims}
\begin{tabular}{lccc}
\hline
Image & RA (J2000) & Dec (J2000) & $\mu$ \\
\hline
Image 1 & $13^h 59^m 48.684^s$ & $62\degr 30\arcmin 48.54\arcsec$ & $12.4 \pm 3$ \\
Image 2 & $13^h 59^m 49.430^s$ & $62\degr 30\arcmin 45.13\arcsec$ & $9.5 \pm 4$\\
Image 3 & $13^h 59^m 54.746^s$ & $62\degr 31\arcmin 5.21\arcsec$ & $2.9 \pm 0.1$\\
\hline
\end{tabular}
\end{table}

Whilst the emission line centroid is extremely well-matched to the best-fit redshift of the nebular emission, the line is weak, highlighting the difficulty of these observations even with long integrations. We perform a number of tests to validate the robustness of the detection. First, we check how often a detection is made when coadding two random spectra from the original (non-coadded) cube. We generate 1000 combinations of two randomly-selected pixels on the map (excluding pixels which lie within one beam of MS1358-arc), then fit a Gaussian profile to the resulting coadded spectrum and compare this to a continuum-only fit. The probability of finding an emission line of equal significance to the target from two randomly-selected positions with a velocity centroid lying within $\pm 150$km\,s$^{-1}$ of the redshift of MS1358-arc and a line width in the range $\sigma = 30 - 250$\,km\,s$^{-1}$ is 0.05\%.

The CO(5--4) emission line we measure is centered at $z = 4.9297 \pm 0.0001$ with a line flux of $0.104 \pm 0.024$\,Jy\,km\,s$^{-1}$ and Full Width at Zero Intensity of FWZI $= 150 \pm 20$\,km\,s$^{-1}$, comparable to the velocity gradient measured across the source by \citetalias{2009MNRAS.400.1121S}. Error bars are obtained by fitting a Gaussian profile to the line and perturbing it in intensity, velocity and width to obtain a $\Delta\,\chi^2 = 1$ error surface, using a Monte Carlo routine with $10^5$ realisations centered on the best fit.

\begin{figure}
\includegraphics[width=84mm]{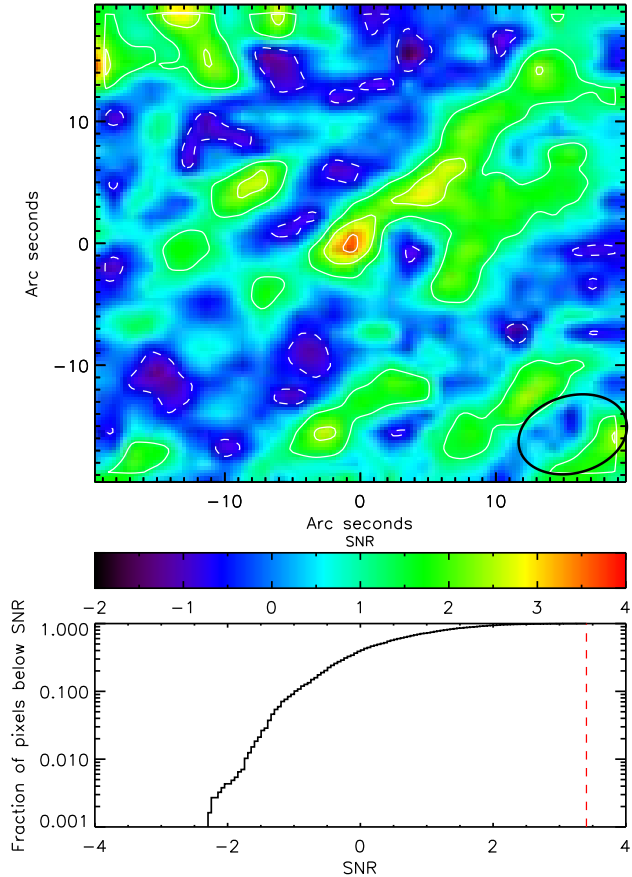}
\caption{\emph{Upper:} The coadded PdBI signal-to-noise channel map, constructed by
  extracting and coadding regions around each of the two images of
  MS1358-arc, convolved spatially with the beam and in the spectral direction with a Gaussian profile with FWHM$ = 100$km\,s$^{-1}$, and then dividing each pixel by its noise. The combined image of the two targets is located in the centre of the map. The solid lines are intensity contours in $\pm 1\sigma$ intervals, with negative contours indicated by dashed lines. The beam
  size is indicated by the black ellipse line in the lower
  right-hand corner.
\emph{Lower:} Cumulative histogram of the signal-to-noise in the map, with the central pixel marked as a red dashed line. We find that just $0.05\%$ of pixels have higher signal-to-noise than the emission seen from the source, suggesting a 3.5$\sigma$ detection.}
\label{fig:coim}
\end{figure}

We also construct a coadded channel map from the PdBI cube by extracting regions centered on the two images and co-adding them. The coadded cube is then spatially convolved with the PdBI beam, and spectrally convolved with a Gaussian with FWHM $= 100$\,km\,s$^{-1}$. We estimate the noise at each spatial position using the off-line spectrum at that position, and divide the signal in the on-line slice of the convolved datacube by this noise map to construct a signal-to-noise map, shown in Figure \ref{fig:coim}. We find that 99.95\% of the pixels have lower signal-to-noise than the target. This is equivalent to a detection significance of $3.5\,\sigma$.

In the following analysis, we treat the line flux as $0.104 \pm 0.024$\,Jy\,km\,s$^{-1}$, but discuss the implications of treating it as an upper limit in Section \ref{sec:conc}.

\section{Results, Analysis and Discussion}
\label{sec:results}

\subsection{CO Dynamics, Luminosity and Molecular Gas Mass}

Using the combined spectrum of Images 1 and 2, we first estimate the line luminosity and gas mass. Following
\citet{2005ARA&A..43..677S}, the line luminosity is
$L'_{\rm{CO(5-4)}} = \left(3.6 \pm 0.8 \right) \times
10^9\,\rm{K\,km\,s^{-1}\,pc}^2$. Adopting a magnification factor of
$\mu = 22 \pm 5$, this indicates an intrinsic $L'_{\rm{CO(5-4)}} = \left( 1.6 \pm
0.5 \right) \times 10^8$\,K\,km\,s$^{-1}$\,pc$^2$.

To convert this CO(5--4) luminosity to a total molecular gas mass, we
must first estimate the corresponding CO(1--0) line luminosity.  We
therefore exploit recent observations of $z=2-4$ galaxies where the
ratio $L'_{\rm{CO(5-4)}}/L'_{\rm{CO(1-0)}}$ has been measured directly, e.g. $L′_{\rm  CO(5-4)}/L'_{\rm CO(1-0)} = 0.35 \pm 0.02$ for SMM 
\,J2135 \citep{2011MNRAS.410.1687D} or $0.32 \pm 0.05$ for a sample of  
SMGs \citep{2012arXiv1205.1511B}. We adopt the latter value and applying  
it to MS1358-arc, we estimate $L'_{\rm CO(1-0)} = \left(5.0 \pm 1.7\right) \times 10^8$\,K\,km\,s$^{−1}$\,pc$^2$. If we instead assumed $L′_{\rm  CO(5-4)}/L'_{\rm CO(1-0)}$ from either the Antannae, 0.2 \citep{2003ApJ...588..243Z}; NGC\,253, 0.7 \citep{2004A&A...427...45B}; or M\,82, 0.7 \citep{2005A&A...438..533W} this estimate would vary by a factor of $2\times$ in either direction and so we adopt this as the likely systematic uncertainty in our estimate.

To derive the molecular gas mass $M_{\rm{gas}}$ from
$L'_{\rm{CO(1-0)}}$, we assume $M_{\rm{gas}} = \alpha
L'_{\rm{CO(1-0)}}$ where $\alpha$ is the coefficient relating the CO
to H$_2$+He gas mass. We can derive an absolute lower limit for $\alpha$
by assuming that the molecular gas is enriched to solar metallicity and
is optically thin to CO radiation. From \citet{2010MNRAS.404..198I}, we have $\alpha \ga 0.7$, which is
consistent with the value $\alpha = 0.8$ applied to local (Ultra)
Luminous InfraRed Galaxies \citep[(U)LIRGs][]{2005ARA&A..43..677S}. However, recent studies of high-$z$ star-forming galaxies have suggested
that $\alpha \sim 2$ may be a more appropriate conversion factor
\citep[e.g.][]{2011MNRAS.410.1687D,2011MNRAS.tmp...46I}. Adopting this
value, we find $M_{\rm{gas}} = 1^{+ 1}_{- 0.6} \times 10^9\, \rm{M}_{\sun}$, where we conservatively estimate a factor of $2\times$ uncertainty in $\alpha$, comparable to the uncertainty in $L′_{\rm  CO(5-4)}/ L'_{\rm CO(1-0)}$.

Combining our gas mass estimate with the stellar mass $M_{\ast} = 7 \pm 2
\times 10^8$\,M$_{\sun}$ derived by \citetalias{2009MNRAS.400.1121S},
we obtain a total baryonic mass of $M_{\rm{baryon}} = M_{\ast} +
M_{\rm{gas}} = 1.7^{+1.0}_{-0.6} \times 10^9 \rm{M}_{\sun}$. This is consistent with the dynamical mass within 2kpc derived by \citetalias{2009MNRAS.400.1121S} of $M_{\rm{dyn}} = 3 \pm 1 \times 10^9 \csc^2(i)$\,M$_{\sun}$, if the inclination is $i \ga 45\degr$.


The gas fraction $f_{\rm{gas}} =
M_{\rm{gas}}/\left(M_{\rm{gas}} + M_{\rm{\ast}}\right)$ is typically $< 10\%$ in local large
spiral galaxies \citep{1991ARA&A..29..581Y}, or $\sim 33\%$ in local
ULIRGs \citep{1997ApJ...478..144S}. Other studies of molecular gas in
high-redshift galaxies are beginning to find a trend for higher gas
fractions at higher redshifts \citep{2010Natur.463..781T, 2011ApJ...730L..19G}, which is in line with expectations from
hydrodynamical simulations \citep{2009MNRAS.399.1773C}. For MS1358-arc, we find
$f_{\rm{gas}} = 0.59^{+0.11}_{-0.06}$, with a systematic uncertainty of $\sim \pm 0.20$ due to $\alpha$ and $L′_{\rm  CO(5-4)}/L'_{\rm CO(1-0)}$.

\begin{figure*}
\includegraphics[width=170mm, angle=0]{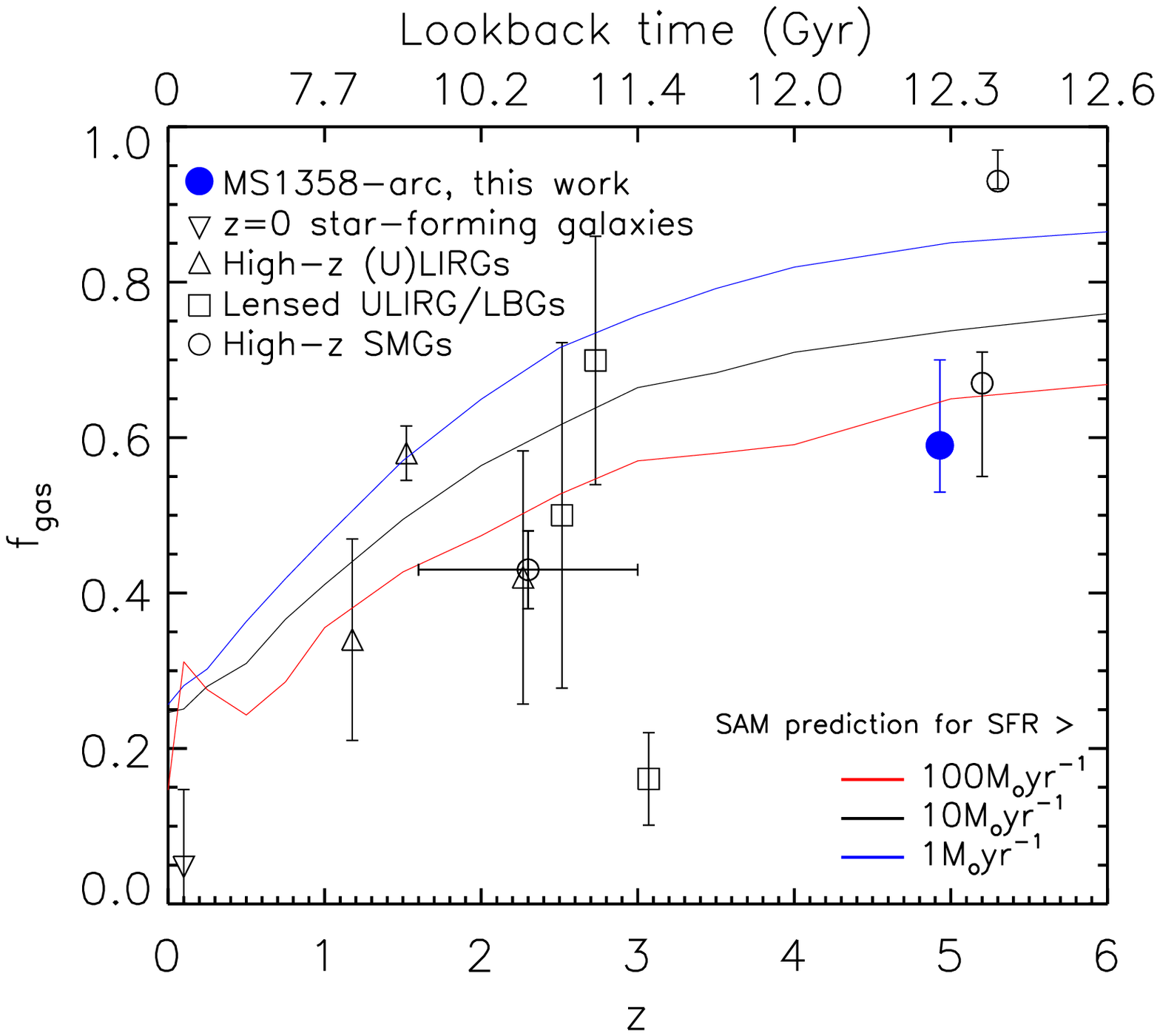}
\caption{Evolution of the molecular gas fraction with redshift in
  star-forming galaxies. We compare MS1358-arc to the $z=0$ sample of
  \protect \citet{2008AJ....136.2782L}, the median gas fractions of the
  high-$z$ LIRGs from \protect \citet{2010ApJ...714L.118D} and the
  median of each redshift bin of the \protect
  \citet{2010Natur.463..781T} (U)LIRG sample. We also use the lensed
  ULIRG from \protect \citet{2005A&A...434..819K}, lensed LBGs from
  \protect \citet{2010ApJ...724L.153R} and high-$z$ SMGs from \protect
  \citet{2012arXiv1205.1511B}, \protect \citet{2010ApJ...720L.131R} and \protect \citet{2012Natur.486..233W}, all converted to $\alpha = 2$ for
  consistency. The lines are the predictions from semi-analytic models
  for galaxies with $SFR > 1,10,100$\,M$_{\sun}\,\rm{yr}^{-1}$ (Lacey
  et al. in prep). From the limited data available, there is apparent
  evolution in the molecular gas fraction with redshift, broadly
  consistent with the semi-analytic models.}
\label{fig:fgasz}
\end{figure*}

In Figure \ref{fig:fgasz}, we compare the gas fraction of MS1358-arc to
samples at lower redshift \citep[see also][]{2011ApJ...730L..19G}. The sample size is
currently too small to draw any firm conclusions, and there are of course
significant selection effects involved. We therefore caution that there are limits to the extent of
any physical interpretation of this result, although it clearly
motivates a uniformly-selected survey of the gas properties of
high-redshift star-forming galaxies.

We can also use the dynamics of the [O{\sc ii}] emission to test whether the molecular gas is colocated with the star formation. In Figure \ref{fig:specs}, we
overplot the redshifts of the two brightest star-forming knots, which are located at $\pm 150$\,km\,s$^{-1}$ from the dynamical centre. Hence, if the CO gas traced the [O{\sc ii}] emission we would expect a FWZI of $\sim 300$km\,s$^{-1}$, higher than the observed line width. This may indicate
that the CO emission is associated with the dynamical centre of the
galaxy rather than being concentrated in the star-forming regions or associated with
the outflowing Ly$\alpha$ and UV-ISM lines
\citepalias{2009MNRAS.400.1121S}.

\subsection{Gas depletion timescales}

\citetalias{2009MNRAS.400.1121S} derived a star formation rate (SFR) of
$42 \pm 8$\,M$_{\sun}\,\rm{yr}^{-1}$ using the
\citet{1998ARA&A..36..189K} conversion from [O{\sc ii}] luminosity to SFR. We note that this conversion is consistent with the empirical correction to this
calibration of \citet{2010MNRAS.405.2594G}, which accounts
for the metallicity dependence of the [O{\sc ii}]-derived SFR.

If MS1358-arc has sustained this SFR continuously, then the time taken
to build up the current stellar mass of $7 \pm 2 \times
10^8$\,M$_{\sun}$ is just $\tau_{\rm{build}} \sim 17 \rm{Myr}$ (consistent with the age and star formation history used to derive the stellar mass). This suggests we may be seeing this galaxy in its first epoch of star formation. If this SFR is maintained at a
constant level, the gas supply of $1 \times 10^9$\,M$_{\sun}$ will be
exhausted in $\sim 24$\,Myr, which would place the galaxy $\sim 40\%$
of the way through its starburst. The star formation `lifetime' of this
galaxy would thus be very short compared to that inferred indirectly for the LBG population at $z =
4-6$, which have implied starburst lifetimes of $\la 500$\,Myr
\citep{2009ApJ...697.1493S}, but is consistent with the lifetime of the starburst-triggered `LBG phase' predicted by semi-analytic models \citep{2011arXiv1105.3731G}.

Finally, we note that the ratio between far-infrared luminosity and molecular gas mass,
$L_{\rm{FIR}}/M_{\rm{H_2}}$, gives the star formation efficiency (SFE),
relating the ongoing star formation to the available molecular
gas. 

\citet{2008MNRAS.384.1611K} obtained an upper limit on the
$850\,\mu$m flux of MS1358-arc of $S_{850} < 4.8$\,mJy which,
accounting for the lensing magnification and using the $S_{850} -
L_{\rm{FIR}}$ calibration of \citet{2003ApJ...597L.113N}, suggests a
far-infrared (FIR) luminosity $L_{\rm{FIR}} < 3.5 \times
10^{11}$\,L$_{\sun}$ (we note that if we assume
that the FIR flux is re-emitted light from dust-obscured star
formation and apply the \citet{1998ARA&A..36..189K} relation to our
[O{\sc ii}]-derived SFR, we obtain $L_{\rm{FIR}} = (2.3 \pm 0.9) \times
10^{11}$\,L$_{\sun}$, which is consistent with this upper limit). We thus obtain $SFE \la 240
\rm{L}_{\sun}$\,M$_{\sun}^{-1}$. For comparison, the median $SFE$ in
local ULIRGS is $49 \pm 6$\,L$_{\sun}$\,M$_{\sun}^{-1}$ when corrected
to $\alpha = 2$, although this rises to $172 \pm 23
\rm{L}_{\sun}$\,M$_{\sun}^{-1}$ if $\alpha=0.8$, which is the factor
commonly applied to ULIRGs \citep{1991ApJ...370..158S}, while SMGs give
$SFE \sim 55^{+20}_{-15} \rm{L}_{\sun}$\,M$_{\sun}^{-1}$ for $\alpha = 2$
\citep{2012arXiv1205.1511B}. Our upper limit is consistent with these values.

\section{Conclusions}
\label{sec:conc}

We have used the IRAM Plateau de Bure Interferometer to search for the
CO(5--4) emission in a $z=4.9296$ galaxy lensed by the foreground
cluster MS1358+62. We were able to observe two images of the galaxy
simultaneously, with a total magnification
factor of $22 \pm 5 \times$. We measure a line flux of $0.104 \pm
0.024$\,Jy\,km\,s$^{-1}$ at the position of the galaxy, yielding a detection at 3.5-4.3$\sigma$. 

The molecular gas shows a relatively narrow velocity range around the systemic redshift, unlike the [O{\sc ii}] emission, which mainly arises from two clumps in the galaxy at $\pm 150$\,km\,s$^{-1}$. This may suggest that the
gas is more centrally concentrated than the star formation.

We derive a total gas mass $M_{\rm{gas}} = 1^{+1}_{-0.6} \times 10^9$\,M$_{\sun}$,
suggesting that this galaxy has a gas fraction $f_{\rm gas} = 0.59^{+0.11}_{-0.06}$, with a systematic uncertainty of $\sim \pm 0.20$ due to $\alpha$ and $L′_{\rm  CO(5-4)}/ L'_{\rm CO(1-0)}$, which is similar to the most gas-rich galaxies at $z \sim 2$. This could imply that gas fractions do not continue to rise significantly beyond $z \sim 2$, though a larger sample is clearly needed to draw any conclusions.

Finally, given the tentative nature of the detection, we consider the implications of treating the measured flux as a 4$\sigma$ upper limit. In this case, the resulting gas mass $M_{\rm{gas}} < 1 \times 10^9$\,M$_{\sun}$ would be lower than expected given the stellar and dynamical masses derived by \citetalias{2009MNRAS.400.1121S}, and the gas fraction would be $f_{\rm{gas}} < 0.6$. This would also imply a gas depletion timescale of $< 24$\,Myr, placing the galaxy more than 40\% of the way through a short starburst.


Our observations highlight the difficulty of measuring gas properties of `representative' star-forming galaxies at $z \sim 5$, an era when many of today's massive galaxies may be undergoing their first major episode of star formation. Probing their basic properties - their stellar and gas content and relation to star formation - will provide important physical quantities which galaxy formation models must reproduce.

\acknowledgments

The authors thank Malcolm Bremer, Seb Oliver and Isaac Roseboom for  
help and useful discussions. RCL  acknowledges a studentship from  
STFC. AMS acknowledges an STFC Advanced Fellowship, and RGB, IRS and  
ACE acknowledge support from STFC.  KEKC and JEG both acknowledge the  
support from NSERC. RAC is supported by the ARC. These observations  
were carried out with the IRAM Plateau de Bure Interferometer. IRAM is  
supported by INSU/CNRS (France), MPG (Germany) and IGN (Spain).



{\it Facilities:} \facility{IRAM:Interferometer}.

\bibliographystyle{apj}





\end{document}